\documentclass[journal]{IEEEtran}
\usepackage{stfloats}
\usepackage{graphicx}
\usepackage{booktabs}
\usepackage{float}
\usepackage{caption}
\usepackage{amsmath}
\usepackage{amssymb,amsfonts}
\usepackage{listings}
\usepackage{url}
\usepackage{algorithm,algorithmicx}
\usepackage{algpseudocode}
\usepackage{bm}
\usepackage[switch]{lineno}

\ifCLASSINFOpdf
\else
\fi
\begin{document}
%
\title{Real-time Magnetometer Disturbance Estimation via Online Nonlinear Programming}
%
%
%

\author{Jin~Wu,~\IEEEmembership{Member,~IEEE}
\thanks{This research was supported by National Natural Science Foundation of China under grant 41604025. ({\emph{Corresponding author: Jin Wu}}).}
\thanks{The author is with Department of Electronic and Computer Engineering, Hong Kong University of Science and Technology, Hong Kong, China. (e-mail: jin\_wu\_uestc@hotmail).}
}

\maketitle

\begin{abstract}
Magnetometer is a significant sensor for integrated navigation. However, it suffers from many kinds of unknown dynamic magnetic disturbances. We study the problem of online estimating such disturbances via a nonlinear optimization aided by intermediate quaternion estimation from inertial fusion. The proposed optimization is constrained by geographical distribution of magnetic field forming a constrained nonlinear programming. The uniqueness of the solution has been verified mathematically and we design an interior-point-based solver for efficient computation on embedded chips. It is claimed that the designed scheme mainly outperforms in dealing with the challenging bias estimation problem under static motion as previous representatives can hardly achieve. Experimental results demonstrate the effectiveness of the proposed scheme on high accuracy, fast response and low computational load.
\end{abstract}


%
\IEEEpeerreviewmaketitle

\section{Introduction}
%
%
%
%
\IEEEPARstart{T}{hree}-axis magnetometer is a crucial component in current mechatronic navigation and control applications \cite{Psiaki1993, Zanchettin2013}. It is frequently employed for autonomous heading determination in robotics, which is an important preliminary for further estimation of velocity and position. \\
\indent In orientation estimation, magnetometer can be easily interfered by outer electromagnetic disturbances generated by hard-iron objects, complicated operating environments and time-varying electric currents from motors and transmission wires \cite{Paull2014, Huq2015}. Extensive studies have been performed to solve the problem of offline magnetometer calibration and alignment to inertial sensors \cite{Zhang2015, Zhang2015-2}. However, offline calibration can not fix the issue of online unknown magnetic disturbance. Among many existing robust attitude estimation algorithms, to deal with unknown magnetic  disturbances, there have been quite a lot using intelligent detection of anormalies along with covariance adaption of sensed magnetic field \cite{Lee2017,Beravs2014}. Adaptive covariance estimation is practical in engineering applications but can not fully eliminate the effect of disturbance imposed on the steady-state results \cite{Allotta2016}. Such method can be treated as a dynamical weighting approach maximumly separating the distorted sensor measurements. Besides, magnetic outlier rejection, although proved to be  feasible in high-end navigation systems comprising fiber-optic gyroscopes (FOGs), may also lead to long-term large heading drift for low-cost sensor arrays. Instead, more robust heading determination originates from estimating real-time magnetic disturbances. Online magnetic disturbance estimation is not a new topic, but has been studied long ago for spacecraft missions \cite{Crassidis2005}. Current results sometimes require the geomagnetic model information and needs long time (hour level) to converge to the simulated true values \cite{Soken2017}. Recently, the concept of online calibration of magnetometer and alignment to inertial sensors has become a novel and leading tool, which can be achieved via Kalman filtering design \cite{Wu2018tcst} for full-parametric real-time calibration including scale factors, misalignments and biases. Simpler methods with less estimation parameters for low-cost sensors have been implemented recently using an extended Kalman filter (EKF) as well \cite{Han2017}.\\
\indent The joint limitation of \cite{Wu2018tcst} and \cite{Han2017} is that they cannot deal with static estimation of magnetic disturbances. Rather, representative motions have to be designed to obtain precise calibration results. In fact, for most cases, with magnetic disturbances within full measurement range, the magnetometer can hardly be magnetized and only the online sensor bias is required for heading compensation, which describes the good nonlinearity within sensor measurement saturation. As such, Fedele et al. proposed an asymptotic observer based on Volterra integral with the aid of angular rate measurements \cite{Fedele2018}. It also remains the shortcoming of mandatory distinctive angular motions and cannot instantly obtain the magnetic disturbance since a long window of historical magnetic sensing values is needed. Sophisticated generalized inverse imposed on the large window also highly increases the computational and storage burdens. Representative motions can only be conveniently applied to those cases when the sensors can be rotated and translated. However, for large-scale unmanned aerial vehicles (UAVs) and autonomous underwater vehicles (AUVs), it is very difficult for engineers to uninstall significant sensors or generate distinctive representative motions.\\
\indent Based on limitations shown above, this paper proposes a novel online magnetic disturbance estimator using nonlinear programming. It mainly solves the problem of static estimation of unknown outer magnetic disturbances and also performs well in dynamic mode. It is mainly motivated by aligning the magnetically distorted attitude quaternion to a better reference one so that the disturbance may be estimated. The proposed solution utilizes an interior-point optimum searcher and has been implemented on embedded micro chips as well.\\
\indent This paper is organized as follows: Section II contains the studied problem and our proposed optimization solution. Static and dynamic experiments are presented in Section III to show the superiority of the developed method. While concluding remarks are drawn in the last Section IV.

\section{Proposed Solution}
\subsection{Problem Formulation}
We denote $\bm{M}^b$ as the measured magnetic field vector in the body frame $b$. The in-run magnetometer measurement is related with its true value $\bm{\hat M}^b$ by an additive unknown dynamic magnetic disturbance $ \bm{b}_{\rm{dyn}}=(b_{m_x}, b_{m_y}, b_{m_z})^T$ and a white-Gaussian noise $\bm{\eta}_{\bm{M}^b}$, such that it can be adequately modeled as follows \cite{Fedele2018}
\begin{equation}\label{m_model}
\bm{M}^b = \bm{\hat M}^b + \bm{b}_{\rm{dyn}} +  \bm{\eta}_{\bm{M}^b}
\end{equation}
The passive determination of $ \bm{b}_{\rm{dyn}}$ can be reduced to estimating the true value $\bm{\hat M}^b$. Note that when exposed to very strong soft-iron magnetic disturbances, (\ref{m_model}) is not fully effective as the scale factors are also distorted. However in this paper, we inherit the model in \cite{Fedele2018} so that it can deal with most cases where hard-iron and not very strong soft-iron disturbances coexist. In the following parts, we are going to do such estimation with the hypothesis that magnetometer measurements reflect the heading angle of the attached object while the heading information can also be acquired from other sensor fusion results e.g. inertial-only or vector-aided algorithms \cite{Wu2014, Wu2018}. First, we estimate the normalization of $\bm{\hat M}^b$. \textbf{We assume that} the employed magnetometer has been pre-calibrated for scale, misalignment and biases before onboard data sampling. This can actually be done by many offline means e.g. \cite{Vasconcelos2011,Gebre-Egziabher2006}, as also described in \cite{Fedele2018}. \textbf{Then it is assumed that} the the vehicle on which magnetometer is mounted moves within a section of local area without very large distance from origin (less than 500Km). In this way, $\left\| \bm{\hat M}^b \right\|$ can be regarded as constants fixed in the global Earth frame due to pre-calibration of magnetometer \cite{Fedele2018}. Then, the ideal magnetometer vector can be restored, from which the magnetic disturbance can be accordingly generated by $\bm{b}_{\rm{dyn}} \approx \bm{M}^b - \bm{\hat M}^b$. When the vehicle runs with large distance, then the magnetometer norm $\left\| \bm{\hat M}^b \right\|$ can be also referenced from the International Geomagnetic Reference Field (IGRF, \cite{thebault2015international}) model provided that the global positioning information is known, such that \cite{Abdelrahman2011}
\begin{equation*}
\begin{small}
\begin{gathered}
\left\| \bm{\hat M}^b \right\|^2 = \hfill \\
\begin{gathered}
  {\left[ {\sum\limits_{n = 1}^k {\sum\limits_{m = 1}^n \begin{gathered}
  {\left( {\frac{a}{{{R_e}}}} \right)^{n + 2}}\frac{{\partial P_n^m(\cos \theta )}}{{\partial \theta }} \times  \hfill \\
  \left( {g_n^m\cos m\varphi  + h_n^m\sin m\varphi } \right) \hfill \\ 
\end{gathered}  } } \right]^2} +  \hfill \\
  {\left[ {\sum\limits_{n = 1}^k {\sum\limits_{m = 1}^n \begin{gathered}
  {\left( {\frac{a}{{{R_e}}}} \right)^{n + 2}}\frac{{mP_n^m(\cos \theta )}}{{\sin \theta }} \times  \hfill \\
  \left( {g_n^m\sin m\varphi  - h_n^m\cos m\varphi } \right) \hfill \\ 
\end{gathered}  } } \right]^2} +  \hfill \\
  {\left[{\sum\limits_{n = 1}^k {\sum\limits_{m = 1}^n \begin{gathered}
  {\left( {\frac{a}{{{R_e}}}} \right)^{n + 2}} (n + 1)P_n^m(\cos \theta ) \times  \hfill \\
  \left( {g_n^m\cos m\varphi  + h_n^m\sin m\varphi } \right) \hfill \\ 
\end{gathered}  } } \right]^2} \hfill \\ 
\end{gathered} 
\end{gathered}
\end{small}
\end{equation*}
with $a$ denoting the altitude with respect to the center of the Earth; $R_e$ the standard equivalent Earth radius; $\theta$ and $\varphi$ co-latitude and longitude angles in $rad$; $P_n^m$ the associated Legendre function of degree $m$ and order $n$; $k$ the approximation order; $g_n^m$ and $h_n^m$ Gaussian coefficients from global satellite geomagnetic measurements which are released in IGRF models announced per five or ten years.

\subsection{The Proposed Nonlinear Programming}
Assuming that at a certain time epoch we have measured the normalized vectors from accelerometer and magnetometer denoted as $\bm{a}^b = (a_x, a_y, a_z)^T$ and $\bm{m}^b = (m_x, m_y, m_z)^T=\bm{\hat M}^b/|| \bm{\hat M}^b||$ respectively, the corresponding unnormalized attitude quaternion ${\bm{\tilde q}} = {\left( {{{\tilde q}_0},{{\tilde q}_1},{{\tilde q}_2},{{\tilde q}_3}} \right)^T}$ can be computed as follows \cite{Wu2018tce}
\begin{equation}\label{quat_accmag}
\begin{small}
\begin{gathered}
  {{\tilde q}_0} =  - {a_y}({m_N} + {m_x}) + {a_x}{m_y} \hfill \\ 
  {{\tilde q}_1} = ({a_z} - 1)({m_N} + {m_x}) + {a_x}({m_D} - {m_z}) \hfill \\
  {{\tilde q}_2} = ({a_z} - 1){m_y} + {a_y}({m_D} - {m_z}) \hfill \\
  {{\tilde q}_3} = {a_z}{m_D} - {a_x}{m_N} - {m_z} \hfill \\
\end{gathered} 
\end{small}
\end{equation}
in which $m_D = a_x  m_x + a_y  m_y + a_z  m_z$, $m_N = \sqrt{1 - m_D^2}$. Now applying 
\begin{equation}\label{norm}
\begin{small}
\begin{gathered}
a_x^2 + a_y^2 + a_z^2 =1 \hfill \\
m_x^2 + m_y^2 + m_z^2 =1 \hfill \\
m_N^2 + m_D^2 = 1
\end{gathered}
\end{small}
\end{equation}
the quaternion norm can be given by
\begin{equation*}
\begin{small}
\left\| {{\bm{\tilde q}}} \right\| = 2\sqrt {{m_N}\left[ {\left( {1 - {a_z}} \right)\left( {{m_N} + {m_x}} \right) - {a_x}\left( {{m_D} - {m_z}} \right)} \right]} 
\end{small}
\end{equation*}
When the accelerometer and magnetometer are accurately aligned, the above quaternion determination owns very good precision for estimating $m_N$ and $m_D$. As magnetic disturbances interfere the system, not only quaternion, but $m_N$ and $m_D$ will be distorted as well. The main motivation provided here is that when magnetic disturbances take place, the corresponding yaw and its rate will vary with heading information from quaternions and their derivatives given by inertial/aided fusion. The inertial/aided fusion can be used for checking rate consensus on yaw. Typically, with the zero-angular-rate update (ZARU, \cite{ruiz2012accurate}) in static mode the quaternion from completely inertial fusion can maintain stable and accurate within short period, which is enough for magnetic disturbance compensation. In this way, by comparing the difference of two quaternions, the magnetic disturbances may be estimated.\\
\indent Let us define the \textbf{magnetic vector restoration problem}: 
\noindent With given accelerometer vector $\bm{a}^b$, find $\bm{m}^b$ to achieve the following minimization
\begin{equation}\label{opt1}
\begin{small}
\begin{gathered}
\mathop {\arg \min }\limits_{{{\bm{m}}^b} \in \mathbb{U}^3}\hfill \\
\left\{ \begin{gathered}
  {\left[ { - {a_y}\left( {{m_N} + {m_x}} \right) + {a_x}{m_y} - \left\| {{\bm{\tilde q}}} \right\|{{\hat q}_0}} \right]^2} +  \hfill \\
  \left[ {\left( {{a_z} - 1} \right)\left( {{m_N} + {m_x}} \right) + {a_x}\left( {{m_D} - {m_z}} \right) - \left\| {{\bm{\tilde q}}} \right\|{{\hat q}_1}} \right]^2 +  \hfill \\
  {\left[ {\left( {{a_z} - 1} \right){m_y} + {a_y}\left( {{m_D} - {m_z}} \right) - \left\| {{\bm{\tilde q}}} \right\|{{\hat q}_2}} \right]^2} +  \hfill \\
  {\left[ {{a_z}{m_D} - {a_x}{m_N} - {m_z} - \left\| {{\bm{\tilde q}}} \right\|{{\hat q}_3}} \right]^2} \hfill \\ 
\end{gathered}  \right\}
\end{gathered}
\end{small}
\end{equation}
provided that $\bm{\hat q} = (\hat q_0, \hat q_1, \hat q_2, \hat q_3)^T$ is the estimated quaternion from inertial/GNSS/visual/Lidar sensors \cite{Fourati2014, Aghili2016} exactly when the magnetometer is distorted and $\mathbb{U}^3$ denotes the set of all real 3-dimensional unitary vectors. \textbf{It is also noted that} the initial alignment of the yaw angle to true north has been performed before sensor fusion to eliminate the effect of magnetic declination. The local magnetic declination angle can also be referenced using local coordinates with IGRF model or even simply interpolate from empirical tables. Normally, the declination angles can be referenced in advance with the rough knowledge of operating position by referencing declination tables from IGRF.\\
\indent Such optimization aligns $\bm{\tilde q}$ to $\bm{\hat q}$ and obtains $\bm{m}^b$. It has two evident advantages \cite{Wu2018tce}: 
\begin{enumerate}
\item The roll and pitch are not affected by the magnetic measurements while accelerometer measurements will not influence the determination of heading. 
\item The quaternion presented here is explicit and owns the simplest form and nonlinearities compared with all the other existing solutions.
\end{enumerate}
 To let $\bm{\tilde q}, \bm{\hat q}$ have the same roll and pitch information, the vector $\bm{a}^b$ here is reconstructed using $\bm{\hat q}$. Then components of $\bm{\hat q}$ representing roll and pitch will only be intermediate variables without affecting the determination of $\bm{m}^b$. That is to say, the minimization (\ref{opt1}) will compute the magnetic vector with all attention in the yaw direction. As the accelerometer/magnetometer is adequate for full attitude estimation, $\bm{\tilde q}$ can be aligned to any quaternion which depicts the feasibility of estimating $\bm{m}^b$ from such minimization.\\
\indent The current problem occurs that $\left\| {{\bm{\tilde q}}} \right\| $ may not always be real during optimization search. Besides, for the studied optimization (\ref{opt1}), the estimated variables have to be bounded so that they would be reasonable physically and geographically. Furthermore, note that for a quaternion $\bm{q}$, both $\bm{q}$ and its negative $\bm{-q}$ represent the same rotation \cite{Bar-Itzhack2000}. While (\ref{quat_accmag}) can not always make sure that successive quaternions from accelerometer and magnetometer measurements are continuous. Based on above limitations, the optimization (\ref{opt1}) is then revised by estimating magnetic vector $\bm{m}^b$ along with the quaternion norm $k=\left\| {{\bm{\tilde q}}} \right\|$. The new programming is given by
\begin{equation}\label{opt2}
\begin{small}
\begin{gathered}
  \mathop {\arg \min }\limits_{\left[ \begin{gathered}
  {{\bm{m}}^b} \hfill \\
  k \hfill \\ 
\end{gathered}  \right] \in {\mathbb{R}^4}} \left\{ \begin{gathered}
  {\left[ { - {a_y}\left( {{m_N} + {m_x}} \right) + {a_x}{m_y} - k{{\hat q}_0}} \right]^2} +  \hfill \\
  \left[ {\left( {{a_z} - 1} \right)\left( {{m_N} + {m_x}} \right) + {a_x}\left( {{m_D} - {m_z}} \right) - k{{\hat q}_1}} \right]^2 +  \hfill \\
  {\left[ {\left( {{a_z} - 1} \right){m_y} + {a_y}\left( {{m_D} - {m_z}} \right) - k{{\hat q}_2}} \right]^2} +  \hfill \\
  {\left[ {{a_z}{m_D} - {a_x}{m_N} - {m_z} - k{{\hat q}_3}} \right]^2} \hfill \\ 
\end{gathered}  \right\} \hfill \\
  \begin{array}{*{20}{c}}
  {s.t.}&{\left\{ \begin{gathered}
\textbf{Inequalities:} \ \begin{gathered}
  \gamma _{{m_D}}^ -  < \left| {{m_D}} \right| < \gamma _{{m_D}}^ +  \hfill \\
  \gamma _k^ -  < \left| k \right| < \gamma _k^ +  \hfill \\
\end{gathered} \hfill \\
 \textbf{Equalities:}\  (\ref{norm})\  \textbf{and}\  m_D = a_x  m_x + a_y  m_y + a_z  m_z \hfill \\
\end{gathered}  \right.} 
\end{array} \hfill \\ 
\end{gathered} 
\end{small}
\end{equation}
where $\gamma _{m_D}^{+}, \gamma _{k}^{+}$ and $\gamma _{m_D}^{-}, \gamma _{k}^{-}$ are upper and lower bounds for the variables $m_D$ and $k$, respectively. The bounds $\gamma _{m_D}^{+}, \gamma _{m_D}^{-}$ are chosen according to the local magnetic dip angle \cite{Shorshi1995} and $k$'s bounds are set based on the following criterion
\begin{equation*}
\begin{gathered}
  \gamma _k^ +  = {\beta _{\max }}\left| {{a_z} - 1} \right| \hfill \\
  \gamma _k^ -  = {\beta _{\min }}\left| {{a_z} - 1} \right| \hfill \\ 
\end{gathered} 
\end{equation*}
where $\beta_{\max}, \beta_{\min} > 0$ are empirical constants for range scaling. Such criterion is according to the fact that $
  \mathop {\lim }\limits_{{a_z} \to 1} \left\| {{\bm{\tilde q}}} \right\| 
   = 0$ while in such condition
\begin{equation*}
\mathop {\lim }\limits_{{a_z} \to 1} {{\tilde q}_0} = \mathop {\lim }\limits_{{a_z} \to 1} {{\tilde q}_1} = \mathop {\lim }\limits_{{a_z} \to 1} {{\tilde q}_2} = \mathop {\lim }\limits_{{a_z} \to 1} {{\tilde q}_3} = 0
\end{equation*}
as well. For cases that $a_z \to 1$, the norm of quaternion will approach to very tiny values, then $\gamma _{k}^{+}, \gamma _{k}^{-} > 0$ are to ensure proper range for values of $k$ guaranteeing non-existence of indefinite limits $0/0$.

\subsection{Uniqueness of Solution}
Let us conduct the variable replacement by ${\bm{\tilde q}} -  k{\bm{\hat q}}= {\bm{Px}}$ where
\begin{equation*}
\begin{small}
\begin{gathered}
  {\bm{x}} = {\left( {{m_N} + {m_x},{m_y},{m_D} - {m_z},k} \right)^T} =(x_0, x_1, x_2, x_3)^T \hfill \\
  {\bm{P}} = \left[ {\begin{array}{*{20}{c}}
  { - {a_y}}&{{a_x}}&0&{ - {{\hat q}_0}} \\ 
  {{a_z} - 1}&0&{{a_x}}&{ - {{\hat q}_1}} \\ 
  0&{{a_z} - 1}&{{a_y}}&{ - {{\hat q}_2}} \\ 
  { - {a_x}}&{ - {a_y}}&{{a_z} + 1}&{ - {{\hat q}_3}} 
\end{array}} \right] \hfill \\ 
\end{gathered} 
\end{small}
\end{equation*}
This indicates that the solution to the system ${\bm{\tilde q}} -  k{\bm{\hat q}}= \bm{0}$ is equivalent to the null space of $\bm{P}$. However, it should be noted that here the null space of $\bm{P}$ is not uniquely a column vector. Instead, it is composed by two perpendicular vectors. Here we would notice that the system of $\bm{x}$
\begin{equation}\label{ms}
\begin{small}
\left\{ {\begin{array}{*{20}{c}}
  {{m_N} + {m_x} = {x_0}} \\ 
  {{m_y} = {x_1}} \\ 
  {{m_D} - {m_z} = {x_2}} 
\end{array}} \right.
\end{small}
\end{equation}
fully depends on the independent programming governed by $k$. From another aspect, the system (\ref{ms}) can be transformed into quadratic form of $m_x, m_y, m_z$, which indicates there are two independent solutions as $m_y$ is already determined. However, not all solutions can meet the requirements of the programming such that $  \gamma _{{m_D}}^ -  < \left| {{m_D}} \right| < \gamma _{{m_D}}^+$. This reflects that the final optimal solution is still constrained by magnetic-field distribution of the geomagnetic model \cite{Shorshi1995}. Then based on such constraint, the solved $m_x, m_y, m_z$ will be unique in practice. 

\subsection{Interior-Point Method}
As the solution to (\ref{opt2}) can be uniquely determined, we introduce the interior-point method for solving the optimization. Let us define the optimization variable in (\ref{opt2}) as $\bm{y}=\left[ (\bm{M}^b)^T, k\right]^T$ and $f(\bm{y})$ denotes the scalar function to be minimized. Then all the constraints are tantamount to the standard form as follows
\begin{equation*}
\begin{small}
\begin{gathered}
  {c_1}(\bm{y}) = {{\left( {\gamma _{{m_D}}^ + } \right)}^2} - m_D^2 > 0 \hfill \\
  {c_2}(\bm{y}) = m_D^2 - {{\left( {\gamma _{{m_D}}^ - } \right)}^2}  > 0 \hfill \\
  {c_3}(\bm{y})  = {{\left( {\gamma _k^ + } \right)}^2} - {k^2} > 0 \hfill \\
  {c_4}(\bm{y})  = {k^2} - {{\left( {\gamma _k^ - } \right)}^2}  > 0 \hfill \\
  {c_5}(\bm{y})  = m_x^2 + m_y^2 + m_z^2 = 1 \hfill \\ 
\end{gathered} 
\end{small}
\end{equation*}
By introducing the barrier parameter $\rho > 0$, the barrier function is defined by \cite{Nocedal1999Numerical}
\begin{equation*}
\begin{small}
{\mathcal{B}}\left(\bm{y}, \rho \right)=f(\bm{y}) - \rho\left\{{\rm{ln}}[c_1(\bm{y})c_2(\bm{y})c_3(\bm{y})c_4(\bm{y})c_5(\bm{y})] \right\}
\end{small}
\end{equation*}
Now let us employ the Lagrangian multiplier $\bm{\lambda} = (\lambda_1, \lambda_2, \lambda_3, \lambda_4, \lambda_5)^T > 0$ which subjects to the Karush-Kuhn-Tucker (KKT) conditions defining the optimality of the nonlinear optimization, such that ${c_i}{\lambda _i} = \rho $ for $i = 1, 2, 3, 4, 5$. Then the gradient to the barrier function is computed by \cite{Nocedal1999Numerical}
\begin{equation*}
\begin{small}
\begin{gathered}
\bm{g}=(g_0, g_1, g_2, g_3)^T = \hfill \\
\nabla f(\bm{y}) - \rho \sum\limits_{i = 1}^5 {\frac{{\nabla {c_i}\left( {\bm{y}} \right)}}{{{c_i}\left( {\bm{y}} \right)}}} = \nabla f(\bm{y}) - \bm{G}^T \bm{\lambda}\hfill
\end{gathered}
\end{small}
\end{equation*}
where $\bm{G} = \nabla [c_1(\bm{y}), c_2(\bm{y}), c_3(\bm{y}), c_4(\bm{y}), c_5(\bm{y})]^T$. Finally the interior-point method seeks proper searching direction $\bm{\pi}=\left(\bm{\pi}_{\bm{y}}^T, \bm{\pi}_{\bm{\lambda}}^T\right)$ by steepest gradient via the following system \cite{Nocedal1999Numerical}
\begin{equation}\label{search}
\begin{small}
\begin{gathered}
  \left\{ {\begin{array}{*{20}{c}}
  {{\nabla ^2}{\mathcal{B}}}&{ - {{\bm{G}}^T}} \\ 
  {D\left( \begin{gathered}
  {\lambda _1} \hfill \\
  \hfill \vdots \hfill \\ 
{\lambda _5} \hfill \\ 
\end{gathered}  \right){\bm{G}}}&{D\left[ \begin{gathered}
  {c_1}({\bm{y}}) \hfill \\
  \hfill \vdots \hfill \\
  {c_5}({\bm{y}}) \hfill \\ 
\end{gathered}  \right]} 
\end{array}} \right\}\bm{\pi} = \left\{ {\begin{array}{*{20}{c}}
  { - \nabla f\left( {\bm{y}} \right) + {{\bm{G}}^T}\bm{\lambda} } \\ 
  {\left[ \begin{gathered}
  \rho  - {c_1}({\bm{y}}){\lambda _1} \hfill \\
  \hfill \vdots \hfill \\
  \rho  - {c_5}({\bm{y}}){\lambda _5} \hfill \\ 
\end{gathered}  \right]} 
\end{array}} \right\} \hfill \\ 
\end{gathered} 
  \end{small}
\end{equation}
where $D$ denotes the diagonal matrix. Choosing a step length $h > 0$, the optimization variables are updated by
\begin{equation*}
\begin{small}
\left(
  {\bm{y}}^T, \bm{\lambda}^T   \right)^T = \left(
  {\bm{y}}^T, \bm{\lambda}^T   \right)^T + h \bm{\pi}
\end{small}
\end{equation*}
As all the Jacobians and Hessians related can be analytically pre-computed, the only computation burden falls into the solution to (\ref{search}) which can be efficiently solved via the singular value decomposition (SVD). In the initialization stage of the interior-point search, the initial value of $\bm{y}$ is chosen as $\bm{y}=(0, 0, 0, 1)^T$. When the optimization progresses over time, the current estimate of $\bm{y, \lambda}$ is calculated based on the previous optimal search for the purpose of improving computational efficiency.

\begin{figure*}[hb]
\setcounter{figure}{4}
  \centering
  \includegraphics[width=1.0\textwidth]{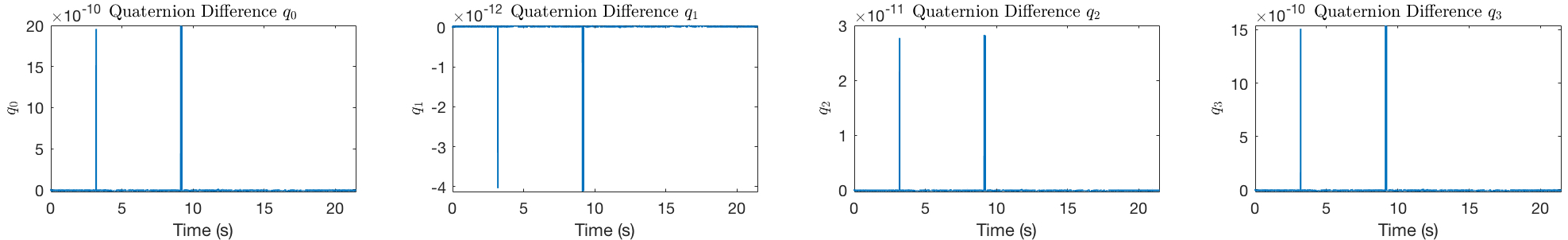}
  \caption{Estimated quaternion differences.}
  \label{fig:q_diff}
%
%
\end{figure*}

\section{Experimental Results}
In this section, we conduct experiments in static and dynamic modes where 'static' one consists no angular and translational motion while dynamic one contains quite distinctive motion and dynamic noises. All the experiments take place in Wuhan, China with the position of latitude and longitude of $E-113^\circ41' \sim 115^\circ 05'$, $N-29^\circ 58' \sim 31^\circ 22'$ respectively. At such position, the theoretical values of $m_N$ and $m_D$ are $m_N = 0.64 \sim 0.69, m_D = -0.77 \sim -0.73$. The standard norm of the magnetic vector after calibration is $\left\| \bm{\hat M}^b \right\|=0.38593\ \rm{Gauss}$. For our proposed method, we set the following parameters:
\begin{enumerate}
\item Inequality constraints: $\gamma_{m_D}^+ = 0.95, \gamma_{m_D}^- = 0.05$, $\beta_{\max}= 10^{4}, \beta_{\min} = 10$.
\item Optimization parameters: initial Lagrangian multiplier: $\bm{\lambda}=(1/5, 1/5, 1/5, 1/5, 1/5)$; barrier parameter: $\rho=10^{-4}$; searching step length: $h=10^{-5}$; maximum iterations: 50; equality constraint tolerance: $10^{-15}$; function value $f(\bm{y})$ tolerance: $10^{-30}$.
\end{enumerate}
The parameters are set very roughly to evaluate the robustness of the proposed algorithm. All the codes related to the proposed method are edited using the C++ programming language while the ALGLIB open-source optimization library of version 3.14.0 has been invoked for solving the proposed nonlinear programming. The codes are compiled via the GNU \texttt{arm-eabi-none-g++-7.0} compiler for program execution on the embedded processing unit STM32H743VIT6. All the run-time performances in this paper are acquired from online computation on STM32H743VIT6 and stored via the SDIO high-speed bus. All the sensors referred to in this paper have underwent rigorous calibration for scale factors, misalignment, biases and etc. The magnetic disturbances are estimated using the proposed method in real-time for all experiments.

\begin{figure}[H]
\setcounter{figure}{0}
  \centering
  \includegraphics[width=0.5\textwidth]{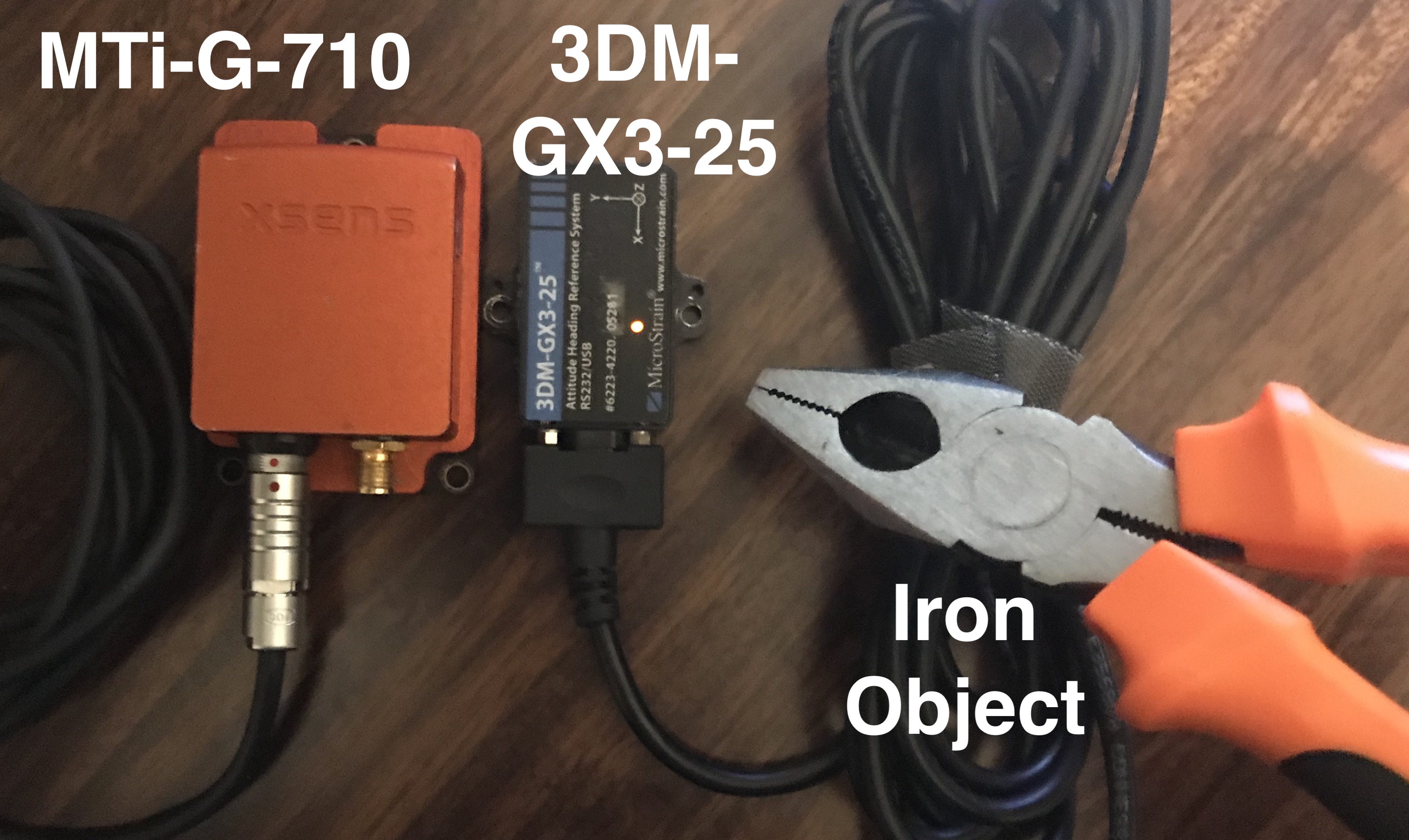}
  \caption{The measured magnetic field vectors from MTi-G-710 and 3DM-GX3-25 are distorted by an iron object.}
  \label{fig:static}
\end{figure}

\subsection{Static Mode}
\begin{figure}[H]
  \centering
  \includegraphics[width=0.5\textwidth]{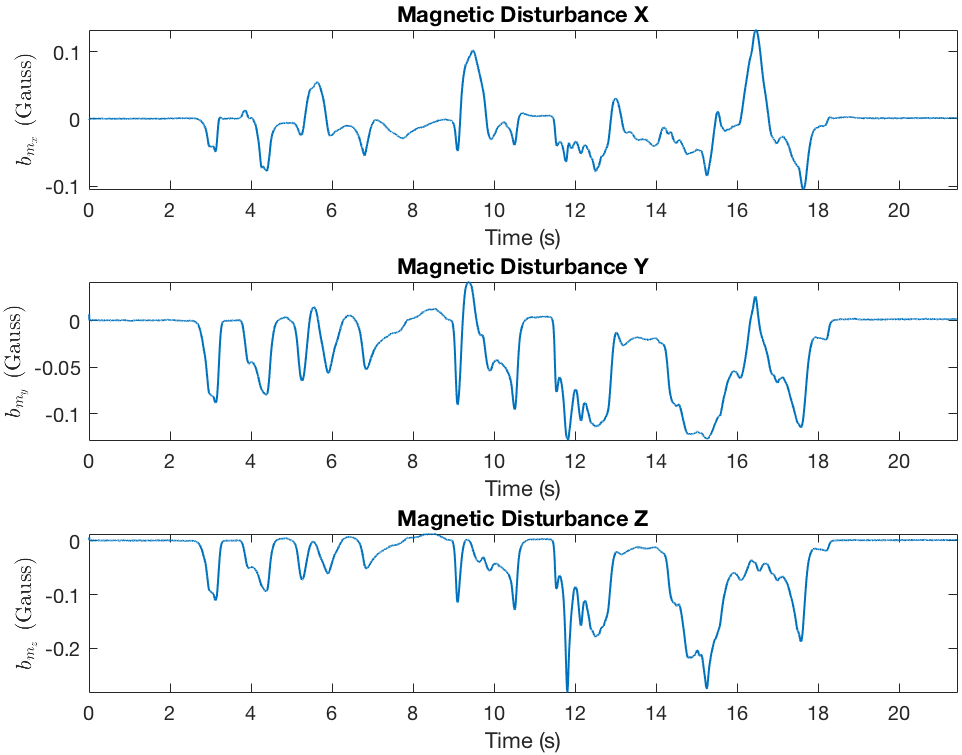}
  \caption{Estimated magnetic disturbances.}
  \label{fig:disturbance}
\end{figure}
\begin{figure}[H]
  \centering
  \includegraphics[width=0.5\textwidth]{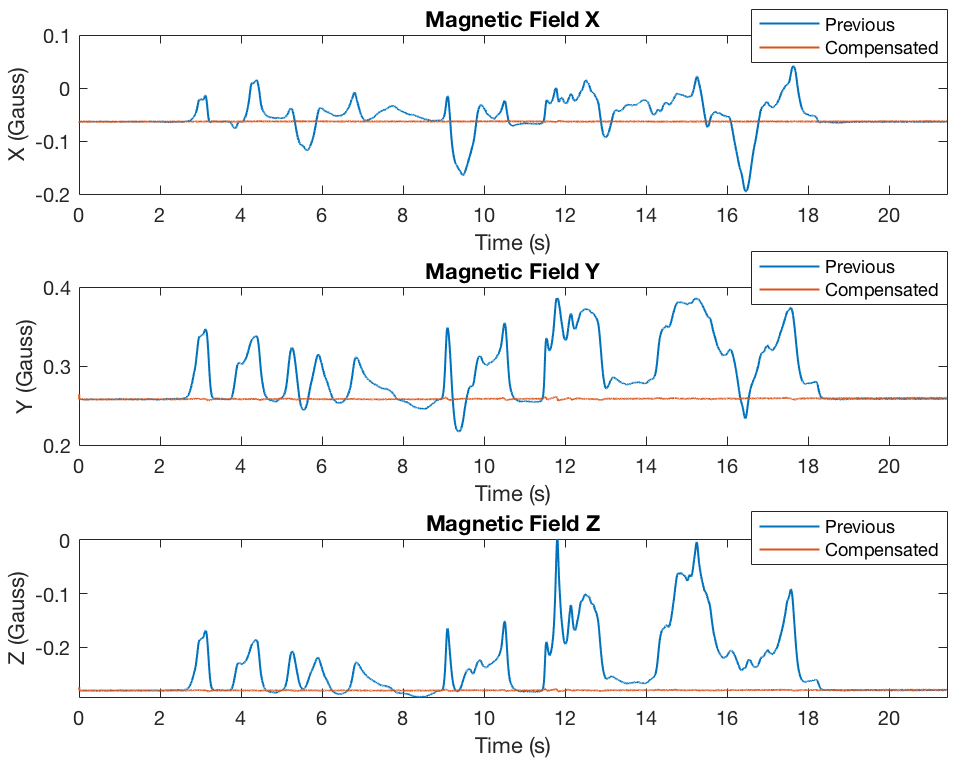}
  \caption{Raw magnetic measurements and compensated ones.}
  \label{fig:compensated_mag}
\end{figure}

In this sub-section, we face a challenging problem of estimating magnetic disturbances in static mode. In such case, the magnetometer stands still and no motion can be acquired to aid the estimation. There is no existing method that can instantly estimate in this situation. We use the hardwares presented in Fig. \ref{fig:static} to illustrate the performances. The Xsens MTi-G-710 integrated navigation product and 3DM-GX3-25 attitude and heading reference system all own internal high-precision 3-axis gyroscope, accelerometer and magnetometer. They are sticked firmly on a testing table and the sensor data of 3DM-GX3-25 has been aligned to that from MTi-G-710. According to Xsens's release notes, MTi-G-710's attitude estimation is immune to magnetic distortion in static mode while 3DM-GX3-25 does not have such functionality. Then based on these characteristics, the raw sensor readings are transmitted from 3DM-GX3-25 in 1000Hz and the readings of magnetometer are downsampled to 50Hz for adequate computational resources for nonlinear programming. The heading results from MTi-G-710 are employed as the source for true value of yaw angles. $\bm{\hat q}$ is obtained by the gyro-accelerometer fusion via a simple complementary filter \cite{Fourati2014}. We use a pair of pincers made of iron to make large magnetic distortion to the sensors. The testing table utilized here does not have any angular and translational motion which make the attached sensors in the fully static mode. With proposed method, the magnetic disturbances are estimated and shown in Fig. \ref{fig:disturbance} while the raw magnetic vectors and compensated ones are depicted in Fig. \ref{fig:compensated_mag}. The quaternion alignment errors $\bm{\tilde q} / k - \bm{\hat q}$ are shown in Fig. \ref{fig:q_diff}.

\begin{figure}[H]
  \centering
  \includegraphics[width=0.5\textwidth]{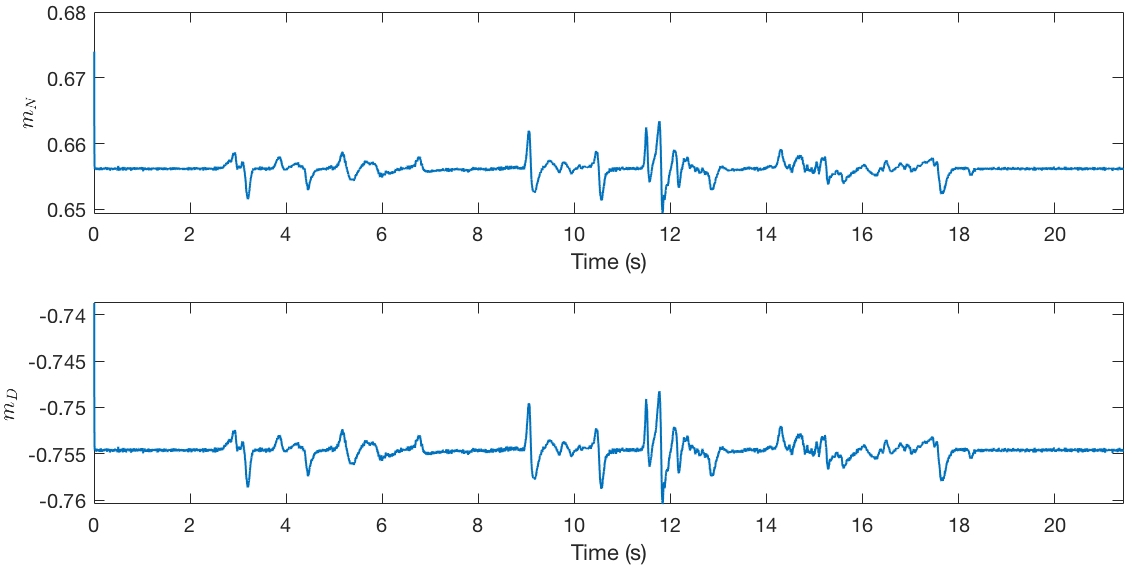}
  \caption{Estimated $m_N, m_D$ values.}
  \label{fig:mN_mD}
\end{figure}

The magnetic disturbances generated here are completely irregular. This can also be indicated by the norms of magnetometer measurements in Fig. \ref{fig:norm}. Moreover, the outer interferences vary instantly with motion of the iron object so previous asymptotically convergent observer can hardly immediately estimate the accurate disturbances. Our method, as a single-point optimizer, acts almost without any delay in such condition and outputs exactly accurate magnetic disturbances. From another side, traditional optimization methods for nonlinear programming are regarded as very slow in execution. However, even running on the embedded processor STM32H743VIT6 with clock speed of 400MHz, the proposed optimization can also accomplished the mission. Although the maximum iteration number has been set to 50, in all the logged run-time results, there is no iteration number over 27 (see Fig. \ref{fig:iter_fval}). And under such times of iteration the core function values can be minimized to a large extent which produces very tiny values not over $5\times 10^{-29}$. Such tiny values coincide with previous quaternion differences in Fig. \ref{fig:q_diff}. For one quaternion with error of $10^{-10}$ level, the magnitude of such errors can be totally ignored in practice. That is to say, the proposed method can estimate both computationally efficient and highly accurate magnetic disturbances.
\begin{figure}[H]
    \centering
  \includegraphics[width=0.5\textwidth]{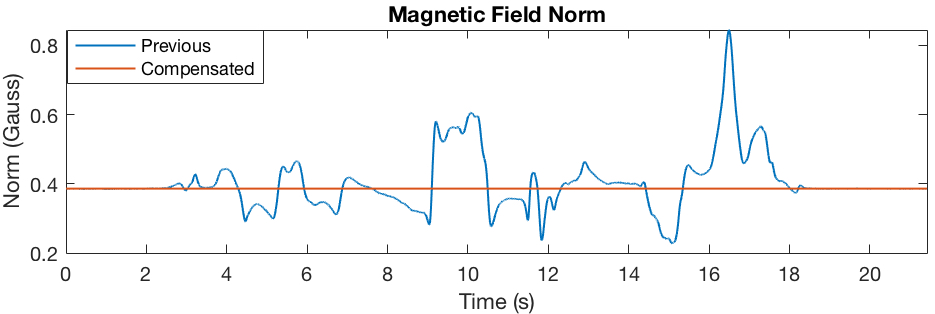}
  \caption{Magnetic norms for previous and compensated magnetic vectors.}
  \label{fig:norm}
\end{figure}
\begin{figure}[H]
\setcounter{figure}{5}
  \centering
  \includegraphics[width=0.5\textwidth]{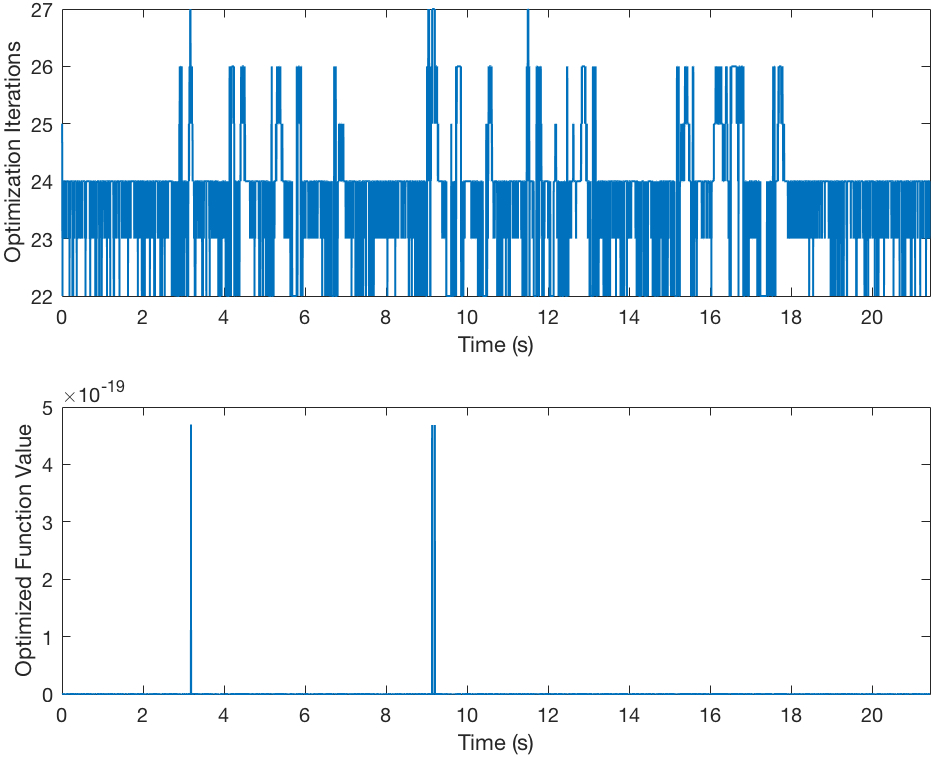}
  \caption{Iteration numbers and minimized function values.}
  \label{fig:iter_fval}
\end{figure}
Also, to verify the physical validity of the proposed method, we present the estimated $m_N$ and $m_D$ values in Fig. \ref{fig:mN_mD}. As described in the beginning of this section, $m_N, m_D$ are in their respective ranges characterized by the geomagnetic model. The presented estimates in Fig. \ref{fig:mN_mD} well fall in such range which reflects its geographical correctness. It is also motivated that, since magnetic field can also be used for positioning in global Earth frame, then applying such estimates of $m_N, m_D$ to previous estimators may also generate a rough information of the latitude, longitude and height. Such information may inversely help engineers to determine the quality of the estimation results.

\subsection{Dynamic Mode}
\begin{figure}[H]
  \centering
  \includegraphics[width=0.5\textwidth]{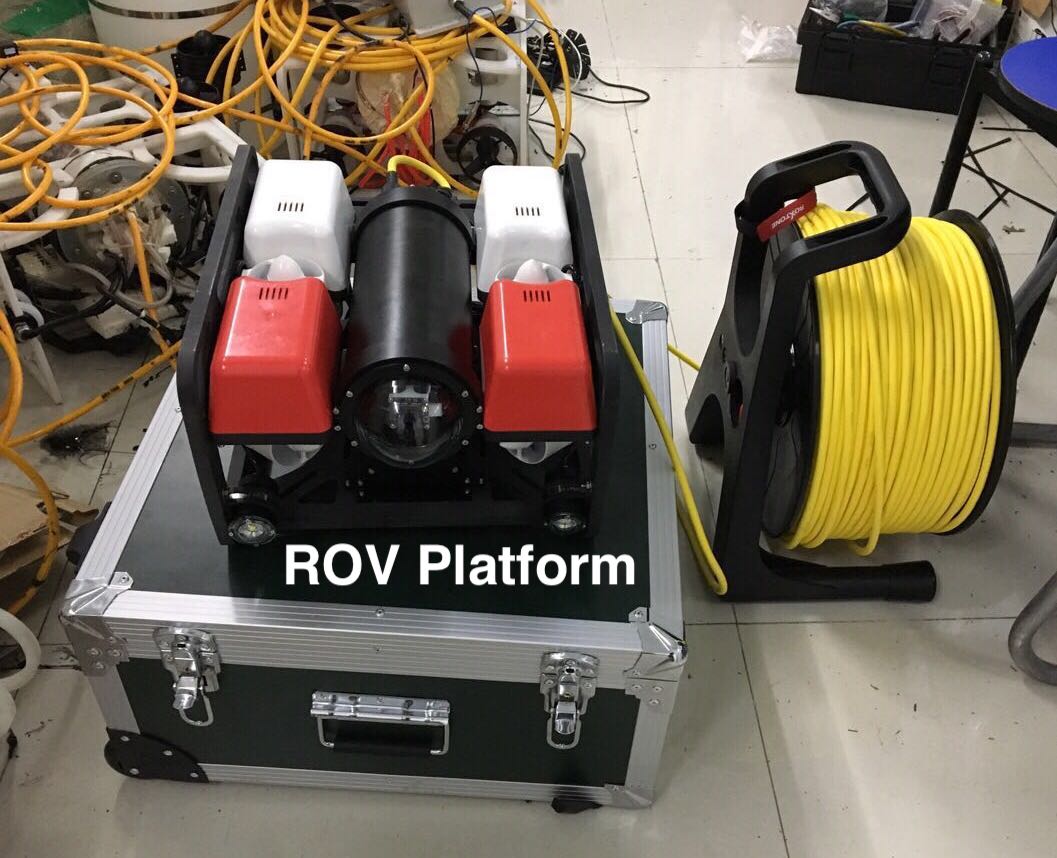}
  \caption{The designed ROV platform.}
  \label{fig:hard1}
\end{figure}
\begin{figure}[H]
%
  \centering
  \includegraphics[width=0.5\textwidth]{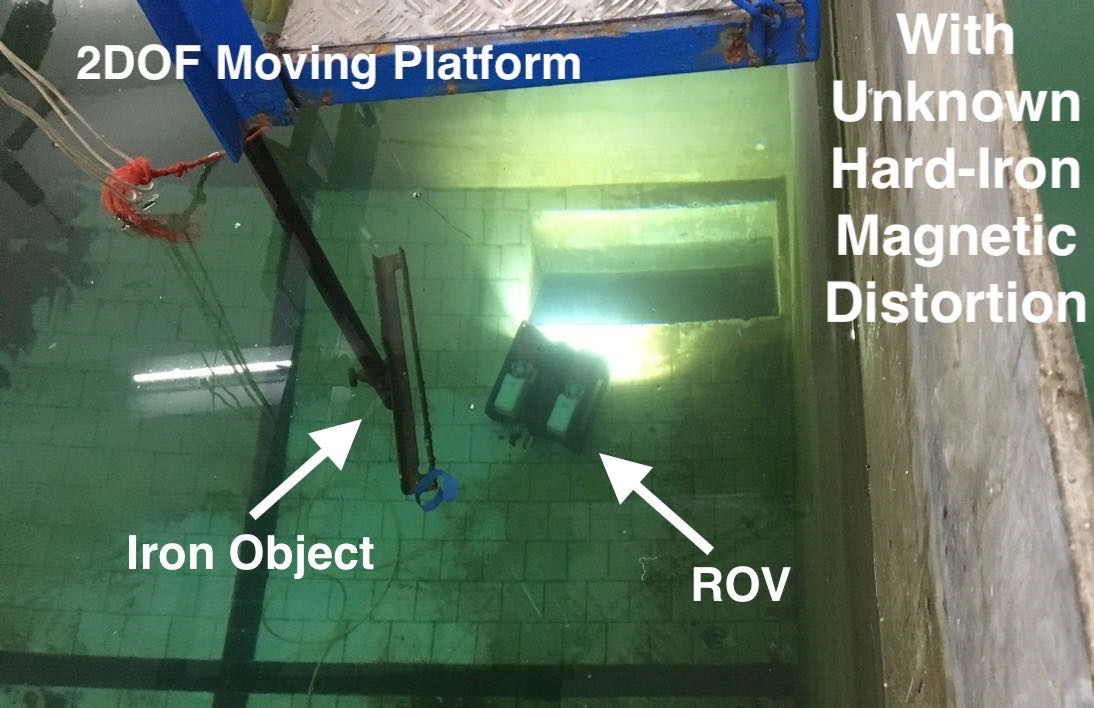}
  \caption{Underwater experiment with iron-magnetic distortion.}
  \label{fig:hard5}
\end{figure}

In this sub-section, dynamic experiments have been carried out on a self-designed remote operated vehicle (ROV, see Fig. \ref{fig:hard1}) in underwater environment. The ROV employs the MTi-G-710 in last sub-section as the navigation sensor. The inertial data is sampled at 400Hz while for magnetometer the frequency is 50Hz. A moving iron-made 2DOF bed is hang over the pool to lead the bottom iron stick into the water (see Fig. \ref{fig:hard5}). The ROV is operated via a ground control system with a joystick and we let it move freely in the water around the iron stick. The proposed scheme is enabled when detected angular rates from accelerometer and magnetometer are significantly less or larger than that from gyroscope (here threshold is set as $10^\circ/hour$. 

\begin{figure}[H]
  \centering
  \includegraphics[width=0.5\textwidth]{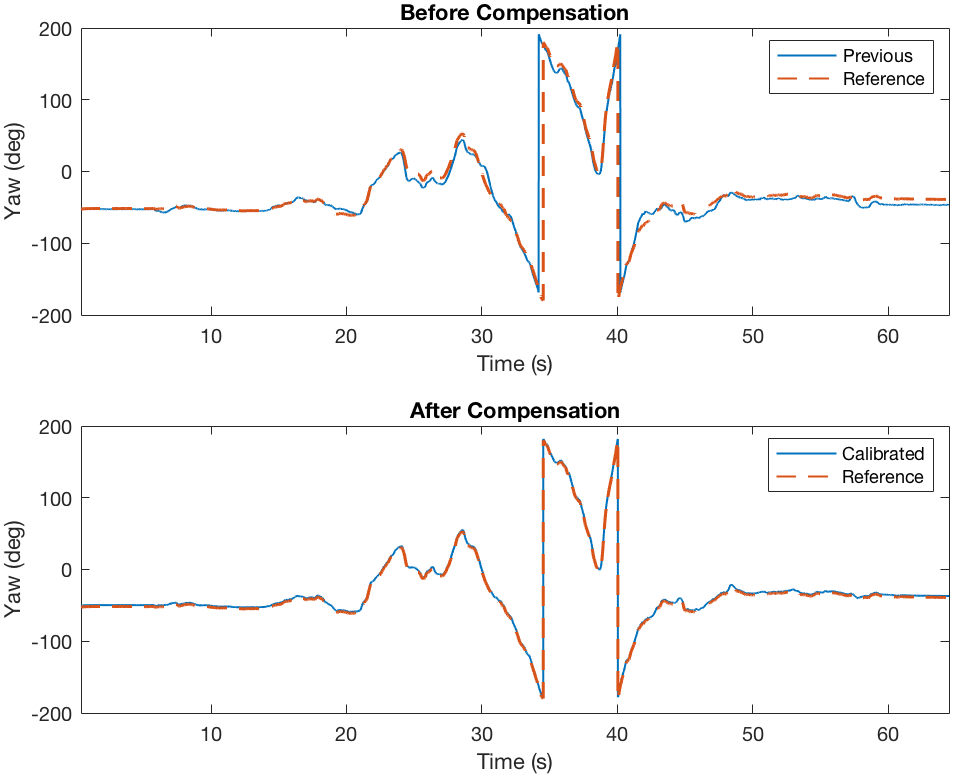}
  \caption{Heading angles before and after compensation.}
  \label{fig:yaw_dyn}
\end{figure}

In such scenario, the magnetometer is distorted both by the iron object and electric currents from thrusters. Such disturbances are generated by motors, which are regarded to be noisy. The proposed method accurately estimates the disturbances and then gives it back to raw sensor readings. Using the orientation method in \cite{Fourati2014}, the heading angles before and after compensation are shown in Fig. \ref{fig:yaw_dyn}. The proposed nonlinear programming can eliminate the magnetic disturbance in a fundamental manner. Therefore the heading determination can be significantly improved. The statistics of accuracy are shown in Table \ref{tab}.
\begin{table}[H]
\centering
\caption{Root Mean Heading Accuracy}\label{tab}
\resizebox{0.35\textwidth}{!}{
\begin{tabular}{ccc}
\toprule
{Before Compensation}&{After Compensation}\\
\midrule
{$12.89265227^ \circ$}&{$1.04376031^ \circ$}\\
\bottomrule
\end{tabular}}
\end{table}
As heading is extremely important for absolute navigation and control in global Earth frame, the proposed method may benefit to related applications in the future.

\section{Conclusion}
In this paper, the magnetic disturbance estimation problem is studied. We propose a novel nonlinear optimization approach to solve such problem by means of the interior-point method. The uniqueness of the solution has been proven via mathematical constraints which ensures the effectiveness of the proposed method. Throughout real-world experiments, it has been validated to be correct, computationally efficient and it owns fast response facing unknown magnetic distortion. Current method can only deal with hard-iron or not very strong soft-iron disturbances, restricted by the model presented in (\ref{m_model}). In the future we may also combine the motor-disturbance model with the proposed work together, forming a complete disturbance identification system \cite{Zhou2016}. Further efforts should be devoted to obtaining more generalized and simplified optimization framework and faster calculation process to achieve better computational performance on low-cost and power-saving applications. It is also noted that the proposed method can only be effective for magnetometer data within full measurement range. Another task for us to accomplish next is to study better algorithm under sensor saturation for more robust estimation performance.


%

%

\section*{Acknowledgment}
This research was supported by National Natural Science Foundation of China under the grant of No. 41604025. We also genuinely thank Prof. Yuanxin Wu from Shanghai Jiao Tong University for his constructive comments.

\ifCLASSOPTIONcaptionsoff
  \newpage
\fi

\begin{IEEEbiography}[{\includegraphics[width=1in,height=1.25in,clip,keepaspectratio]{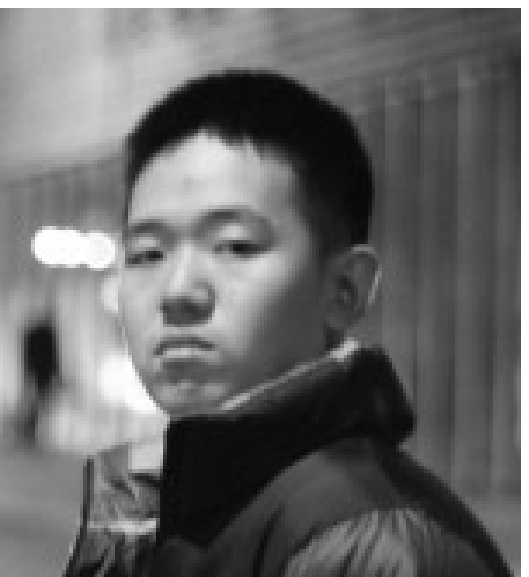}}]{Jin Wu}
(SM'15, M'17) was born in May, 1994 in Zhengjiang, Jiangsu, China. He received the B.S. degree from University of Electronic Science and Technology of China, Chengdu, China. He has been a research assistant in Department of Electronic and Computer Engineering, Hong Kong University of Science and Technology, Hong Kong, China since 2018. His research interests include inertial navigation, optimal filtering, control theory and robot vision. He is a member of IEEE.
\end{IEEEbiography}

\end{document}